\documentclass[pra,bibnotes,twocolumn]{revtex4}
\usepackage{graphicx}
\begin{document}
\draft
\def\ds{\displaystyle}
\title{Self-Accelerating Matter Waves }
\author{C. Yuce}
\address{Department of Physics, Anadolu University, Turkey }
\email{cyuce@anadolu.edu.tr}
\date{\today}
\begin{abstract}
The free particle
Schrodinger equation admits a
non-trivial  self-accelerating Airy wave packet solution. Recently, the Airy beams that freely accelerate in space was
experimentally realized in photonics community. Here we present
self-accelerating waves for the Bose-Einstein condensate in a time
dependent harmonic oscillator potential. We show that parity and time reversal symmetries
for self accelerating waves are spontaneously broken. 
\end{abstract}
\maketitle

\section{Introduction}

In 1979, Berry and Balazs theoretically showed that the
Schrodinger equation describing a free particle admits a
non-trivial Airy wave packet solution \cite{berry}. This free particle wave packet
is unique in the sense that it accelerates. Furthermore, the  Airy wave packet  doesn't spread out as it accelerates. The Airy wave packet is also called self-accelerating wave packet since it accelerates in the absence of an external potential. The accelerating behavior is not consistent with the Ehrenfest
theorem, which describes the motion of the center of mass of the
wave packet. The reason of this inconsistency is the non-integrability of the Airy function. The Berry and Balazs' paper initiated extensive theoretical investigation on such waves \cite{gen1,gen2,gen3,gen4,gen6,gen7,gen8}. The
self-accelerating Airy wave packet was also experimentally realized within
the context of optics three decades after its theoretical
prediction \cite{deney1,deney2,deney3,deney4,deney5,ek}. Since
then, self-accelerating wave packets have stimulated growing
research interest \cite{11,112,113,114,12,13,14,15}. Soon after the first
experimental realization of the self-accelerating Airy beams in
optics, the same types of beams using free electrons instead of
light was generated \cite{electron}. Using a nanoscale hologram
technique, the Airy wave packet for electrons is obtained by
diffraction of electrons. The self-accelerating and non-spreading
solution is not restricted to the linear Schrodinger equation and
nonlinear generalization was considered by some authors \cite{deney4,14}.\\
The ultracold systems are used to study the effects in quantum and statistical physics both theoretically and experimentally. Experimental achievements and theoretical studies of Bose Einstein condensates
(BEC) of weakly interacting atoms have stimulated intensive interest in
the field of atomic matter waves \cite{mw1,mw2}. Recently, self accelerating matter
waves are considered in the absence of external potential \cite{mw3}. In experiments of BEC, harmonic potential is mostly used to trap atoms. Therefore self accelerating matter waves in the presence of external harmonic potential is worth studying.  In this
paper, we find and examine self accelerating wave packets for ultracold atoms in a time dependent harmonic trap. We derive a formula for self-acceleration of such waves and show that
the self-acceleration depends on the initial form of the wave packet.

\section{Self Accelerating Airy Waves}

\begin{figure}[t]
\begin{center}
\centering
\includegraphics[width=6cm]{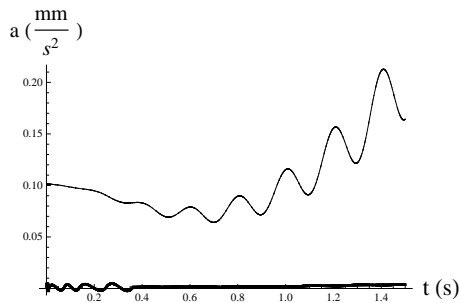}
\caption{ The accelerations of wave packets in units of $mm/s^2$ when $\ds{\omega^2=-2(1+0.2\cos(10\pi t))}$. We take $\ds{a_0=0.1mm/s^2}$ and $x_c(0)=0.5{\mu}m$. The thick curve is plotted according to the Ehrenfest's theorem.  The thin curve is the acceleration for the self-accelerating Airy wave.}
\label{xcddot1}
\end{center}
\end{figure}

In the present study, we consider ultracold atoms in a time dependent trap, which can be realizable experimentally \cite{timettrap}. The physics of ultracold gases in a time dependent trap was theoretically investigated more than a decade ago \cite{theortime1,theortime2}. Another physical freedom on the study of ultracold gases is the temporal tunability of nonlinear interaction. Some experiments on ultracold experiments have demonstrated that tuning of the nonlinear interaction strength can be
achieved by applying an external magnetic field, known as the Feshbach resonance \cite{fesch}. With these experimental degrees of freedom, it is possible to study self-accelerating waves for ultracold atoms. The physics of ultracold atoms is well described by the Gross Pitaevskii (GP) equation \cite{hakem}. The 1-D GP equation for a BEC with time dependent nonlinear interaction strength in a time dependent
harmonic trap reads
\begin{equation}\label{GPjksakdaps}
i\hbar   \frac{\partial \psi }{\partial t}=\left(-\frac{\hbar^2}{2m}  \frac{\partial^2}{\partial  x^2}+
\frac{m\omega^2(t)}{2}  x^2+g(t)|\psi|^2 \right)\psi
\end{equation}
where $\omega(t)$ and $g(t)$ are the time dependent angular frequency and nonlinear interaction strength, respectively. We assume that the nonlinear interaction strength is positive for all time. In the non-interacting limit, the system obeys the Ehrenfest's theorem provided that the wave function is square integrable. According to the Ehrenfest's theorem, the acceleration of wave packet can be found using the equation $\ds{
\frac{d^2<x>}{dt^2}+  \omega^2(t) <x>=0 }$. If either the wave packet is initially displaced from the origin, i.e.,  $\ds{<x(0)>\neq0}$ or the initial velocity of the wave packet is different from zero, i.e.  $\ds{<\dot{x}(0)>=0}$, then the wave packet accelerates. Although the additional presence of the nonlinear interaction changes the density profile, it has nothing to do with the acceleration of the wave packet. The Ehrenfest's theorem shows that the expectation value obeys the classical dynamical laws. Therefore the acceleration of the wave packet matches the acceleration of the classical particle moving in one dimension under the influence of harmonic potential. This statement is true only when the wave packet is square integrable. Below we will find non-integrable wave packet solution of  the GP equation. In other words, we are looking for a wave packet that doesn't obey the Ehrenfest's theorem.  \\
To get self-accelerating wave packet
solution, let us first rewrite the GP equation in an
accelerating frame,
\begin{equation}
x^{\prime }=\frac{x-x_c(t)}{L(t)}
\end{equation}
where  the time dependent function
$\ds{x_c(t)}$ describes translation and $L(t)$ is a time dependent dimensionless scale
factor to be determined later. More precisely, we will see that the width of
the wave packet changes according to $L(t)$. Initially, we take
$L(0)=1$. Under this coordinate transformation, the time
derivative operator transforms as
$\ds{\partial_t\rightarrow\partial_t-(\dot{L}~x^{\prime }+\dot{x_c})/L~
\partial_{x^{\prime}}}$, where dot denotes time derivation.\\
In the accelerating frame, we will seek
the solution of the form 
\begin{equation}\label{uffbea}
\psi(x^{\prime},t)=\frac{1}{\sqrt{L}}~
e^{i\Lambda(x^{\prime},t)} ~\psi({x^{\prime}}) ~,
\end{equation}
where the position dependent phase reads
$\ds{\Lambda(x^{\prime},t)=\frac{m}{\hbar}\left(\alpha
{x^{\prime}}+ \frac{\beta}{2}{x^{\prime}}^2+S\right)}$
and the time dependent functions are given by
$\ds{\alpha(t)=L\dot{x_c}}$, $\ds{\beta(t)= L\dot{L}}$ and
$\ds{\dot{S}(t)=\frac{1}{2} \dot{x_c}^2-\frac{\omega^2}{2}
x_c^2}$. Substitute these transformations into the GP equation and
assume that the following equations are satisfied
\begin{eqnarray}\label{udenklemiyeni}
\ddot{L}+\omega^2(t) ~L&=&0,\\\label{udenklemi}
\ddot{x}_c+\omega^2(t)~x_c&=&\frac{a_0}{L^3},
\end{eqnarray}
where $\ds{a_0>0}$ is an arbitrary
constant in units of acceleration that can be experimentally
manipulated as we shall see below. \\
Suppose the nonlinear interaction changes according to $\ds{g(t)=g_0/L(t) }$, where $\ds{g_0>0}$ is a constant. Then the time dependent GP equation is transformed to the time independent second Painleve equation \cite{21}
\begin{equation}\label{GP3dsksl}
\left(-\frac{d^2}{d{x^{\prime} }^2}+
\frac{2m^2a_0}{\hbar^2}   x^{\prime } +  G_0 |\psi|^2 \right)\psi=0.
\end{equation}
where $\ds{G_0=2m/\hbar^2 g_0}$.\\
One can find a non-integrable stationary solution of this equation. By transforming backwards, the wave packet in the lab. frame can be obtained. However, such a non-integrable solution can't be used to find the acceleration of the wave packet, $\ds{\ddot{ <x>}}$. Fortunately, the equations (\ref{udenklemiyeni},\ref{udenklemi}) describe time dependent width and acceleration of the original wave packet. Before getting stationary solutions of the second Painleve equation, let us discuss these equations in more detail. The equation for $L(t)$ implies
that spreading of the wave-packet is uniquely determined by the
external time dependent harmonic potential. The wave packet
remains non-dispersive in the absence of external potential,
$\omega=0$. Depending on the angular frequency $\omega(t)$, one
can also obtain breathing or expanding/contracting wave packet
solution. The equation (\ref{udenklemi}) implies that the motion
of the wave packet depends not only on the external harmonic potential but
also on the scale factor $\ds{L(t)}$ and the constant $\ds{a_0}$. The constant $\ds{a_0}$ plays a vital role for self accelerating waves since wave packets follow the trajectory of the classical motion in the limit $\ds{a_0=0}$. The self-acceleration depends also on $L(t)$, which is determined by the external potential. The presence of external potential expands/contracts the self accelerating wave and the  expansion/contraction contributes acceleration. It is interesting to note that an initial phase of the wave function plays a role on the self-acceleration. As can be seen from the equation  (\ref{udenklemiyeni}), different initial conditions $\dot{L}(0)$ lead to different scale functions $L(t)$ and this contributes self acceleration according to (\ref{udenklemi}). In other words, imprinting the phase on the initial wave function, i.e., $\ds{\beta(t=0)\neq0}$, changes the self-acceleration of the wave packet. \\
Let us now discuss the solutions of the equations (\ref{udenklemiyeni},\ref{udenklemi}) for some certain cases. The self accelerating solution obtained by Berry and Balasz is recovered when $\omega(t)=0$ \cite{berry}. In this case,
the scale function reads $\ds{L(t)=1+u t}$, where the constant
$\ds{u}$ measures how fast the width of the wave packet changes
(Berry and Balasz considered only the special case with $u=0$). If
$\ds{u=0}$, which is possible by preparing the initial wave
function such that $\beta(0)$ vanishes, the intensity profile of
the wave packet remains invariant while it experiences constant
acceleration. In this case, $\ds{x_c(t)=v_0t+1/2a_0t^2}$, where
$v_0$ is the initial velocity and acceleration matches the
constant $\ds{a_0}$. Observe that acceleration is not due to the
external force. If $\ds{u{\neq}0}$, then the acceleration changes
in time and accelerating wave packet expands/shrinks
ballistically depending on the sign of $\ds{u}$.\\
Consider next the constant harmonic potential, $\ds{\omega^2(t)=\omega_0^2}$. In this case, no
physically acceptable self accelerating solution exists since
$\ds{x^{\prime}}$ becomes singular when the scale function
$\ds{L(t)=\cos(\omega_0t)}$ equals to zero. If the harmonic potential is inverted, $\ds{\omega^2(t)=-\omega_0^2}$ \cite{inv}, the self accelerating wave packet expands or contracts exponentially $\ds{   L(t)=e^{\mp \omega_0 t}   }$ depending on the initial condition $\ds{\beta(0)=\mp\omega_0 t}$.\\
Consider now time dependent harmonic potential. An interesting case is
the non-accelerating wave packet solution in the presence of time
dependent potential. Let us set $\ddot{x}_c=0$ in the coupled
equations (\ref{udenklemiyeni},\ref{udenklemi}). We find that the
acceleration of the wave packet is zero if
$\ds{\omega^2(t)=\frac{2v_0^2}{9(x_0+v_0 t)^2}}$, where $v_0^2=9/2~
a_0$ and $x_0$ is an arbitrary constant. This result contradicts with the
Ehfenfest's theorem, which states that the acceleration of the
moving wavepacket is zero only for the
free particle case.\\
Instead of fixing the acceleration, one can also start with a pre-determined scale factor.
As an example, one can study breathing wave packet by assuming
$\ds{L(t)=1+\epsilon \sin(\Omega t)}$, where $\ds{\epsilon}$ is a
small parameter and the constant $\Omega$ is the angular frequency. We
numerically find that acceleration oscillates between positive and
negative values with growing consecutive peaks. Finally, let us study the evolution of wave packets for a given trap
frequency. As an example, suppose $\omega^2$  changes periodically as $\ds{\omega^2=-\omega_0^2 (1+\epsilon\cos({\Omega} t))}$. We solve the equations
(\ref{udenklemiyeni},\ref{udenklemi}) with the initial conditions
$\ds{x_c(0)=0.5 {\mu}m,\dot{x}_c(0)=\dot{L} (0)=0}$. We take $\ds{a_0=0.1mm/s^2}$, $\epsilon=0.2$, $\omega^2=2~Hz$ and $\Omega=10\pi ~Hz$. In the Fig-\ref{xcddot1}, we compare the accelerations for the non-integrable self-accelerating wave packet and any integrable wave packet.  The Ehrenfest's theorem states that acceleration for integrable wave packets under such a time dependent trap oscillates  non-periodically between negative and positive values with small amplitudes. However, the acceleration of our self-accelerating wave packet oscillates with growing peaks.\\
Let us find the stationary solution of the equation (\ref{GP3dsksl}). In the noninteracting limit, $G_0=0$, an exact analytic solution is available. It is given by
\begin{equation}\label{exactp2}
 \psi(x^{\prime})={\lambda}~Ai((\frac{2m^2a_0}{\hbar^2})^{\frac{1}{3}}x^{\prime})~,
\end{equation}
where $\ds{\lambda}$ is a constant and $Ai(x)$ is the Airy
function. The Airy function is not square integrable and decays exponentially for large positive $\ds{x^{\prime}}$ but
oscillates for large negative $\ds{x^{\prime}}$. One can see that the constant
$\ds{(\frac{2m^2a_0}{\hbar^2})^{\frac{1}{3}}}$ denotes the inverse width
of the main lobe of the initial wave function. Hence we conclude that the constant $\ds{a_0}$
is practically determined by the width of the main
lobe of the Airy function. Therefore acceleration of the Airy wave can be changed by varying the initial width of the Airy function. In this paper, we show that self accelerating waves have unique feature that the acceleration depends on both the external potential and the form of the wave packet.\\
The Airy wave is a stationary solution in the accelerating frame.
Transforming backward yields the accelerating solution in the lab.
frame.
\begin{eqnarray}\label{ufegdsa}
\psi(x,t)=\exp{\left( i\frac{m}{\hbar}(\dot{x}_c {(x-x_c)}+
\frac{\dot{L}}{2L}
{(x-x_c)}^2+S)     \right)}\nonumber\\
\times ~\frac{\lambda}{\sqrt{L}}
~Ai\left((\frac{2m^2a_0}{\hbar^2})^{\frac{1}{3}}~\frac{
x-x_c}{L}\right) ~.~~~~
\end{eqnarray}
This is the exact analytical solution in the noninteracting limit. The wave function is not square integrable and does not obey the Ehrenfest's theorem since the center of mass of the non-integrable Airy
function cannot be defined. To this end, let us mention symmetry properties of self
accelerating waves. The corresponding Hamiltonian for our system remains invariant under both the
parity $\ds{\mathcal{P}}$ operation and time reversal
$\ds{\mathcal{T}}$ operation provided that
$\omega(-t)=\mp\omega(t)$. However, the Airy function is not an
even or odd function of position. Moreover, the time dependent function $x_c(t)$ and the phase
$\Lambda(t)$ are not in general symmetric under time reversal.  One can see from the solution (\ref{ufegdsa}) that $\ds{\mathcal{P}}$ and $\ds{\mathcal{T}}$
symmetries are spontaneously broken for self-accelerating waves.\\
\begin{figure}[t]
\label{cedot2}
\includegraphics[width=7cm]{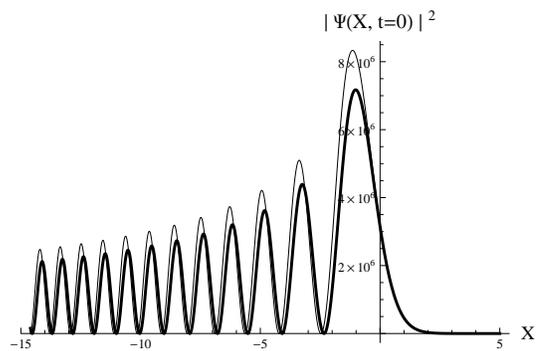}
\caption{ The initial densities for self-accelerating waves when $\lambda=5\times10^3$. The thin and thick curves are for $\ds{G_0=500~m^{-1}}$ (a typical order of magnitude for ultracold atoms in 1D) and $\ds{G_0=0}$, respectively. The densities are plotted with respect to the scaled coordinate $\ds{X=(2m^2a_0/\hbar^2)^{1/3}  x^{\prime}}$, where $(2m^2a_0/\hbar^2)^{1/3}=10^5~m^{-1}$.}
\end{figure}
So far we have considered the non-interacting limit where an exact analytical solution is available. Let us now study self-accelerating waves for the interacting case. It is worth saying that the relations (5,6) are not affected by the nonlinear interaction. Instead, the nonlinear interaction changes the density profile of the self-accelerating wave. We now aim at finding the solution of (\ref{GP3dsksl}). Since no exact solution is available, we solve it numerically. Let us first set the boundary conditions. The nonlinear interaction is dominant around
the main lobe and gets weaker away from the main lobe. Since the density
and consequently the nonlinear interaction go to zero as $\ds{x^{\prime}\rightarrow\infty } $, the
solution of  (\ref{GP3dsksl}) asymptotically coincides with
the Airy function at large positive $\ds{x^{\prime}}$. Therefore we set the boundary condition
$\psi  \sim \lambda Ai( (2m^2a_0/\hbar^2)^{1/3} x^{\prime}) $ and $\partial_{x^{\prime}} \psi  \sim\lambda \partial_{x^{\prime}}  Ai(  (2m^2a_0/\hbar^2)^{1/3}   x^{\prime})$ at large $\ds{x^{\prime}}$ to solve the second Painleve equation numerically. Before going further, we remark that the constant
$\lambda$ is a nontrivial degree of freedom in the interacting case. Increasing $G_0$ at fixed $a_0$ and $\lambda$ increases the peak value of the main lobe slightly. If $G_0$ is very close to a critical value, $G_{c}$, then the peak value changes drastically.  At exactly the critical value, $\ds{G_0=G_{c } = \lambda^{-2} (\frac{2m^2a_0}{\hbar^2})^{3/2} }$, the peak value goes to infinity. Therefore the solution of the equation (\ref{GP3dsksl}) is not bounded if $G_0$ exceeds the critical value, $G_{c}$. \\
The Fig-2 plots the density of the self-accelerating solution at $\lambda=5\times10^3$ for $G_0=500m^{-1}$ (thin curve) and $G_0=0$ (thick curve). As expected, the effect
of the nonlinearity can be mostly seen on the main lobe since the nonlinear
interaction is weak for large $|x^{\prime}|$. The peak value of the main lobe is increased by positive nonlinear interactions and the wave
packet is extended with respect to noninteracting
case. Furthermore, the main lope is
shifted to the left when compared to the linear one. Note that no normalization is made on these solutions since they are not integrable. To this end, we would like to emphasize that time evolution of the nonlinear solution is exactly the same as the Airy wave packet (8). In other words, the solution  given in the Fig.2 translates while expanding/contracting in time according to (2), where $x_c(t)$ and $L(t)$ are the same for both noninteracting and interacting cases.\\
We have shown that non-integrable wave packets lead to
surprising physical results. However, realizing non-integrable
wave function is impossible in practice. Introducing an
exponential aperture function is one possible way for the physical
realization of such waves $\ds{\psi\rightarrow\Psi=e^{{\epsilon}X}\psi}$,
where the decay factor $\ds{\epsilon>0}$ is a positive parameter
to guarantee the convergence of the function on the negative
branch \cite{deney1}. The Fourier transform of the truncated Airy wave packet is
proportional to $\ds{e^{-\epsilon k^2} e^{i(\hbar^2/6 m^2
a_0)k^3}}$ \cite{ek}. In photonics community, the Airy wave was experimentally obtained by passing
a Gaussian beam through a phase mask that adds
a cubic phase modulation \cite{deney1}. Since the wave packet is truncated, acceleration
maintains for a finite distance only. The decay factor
$\ds{\epsilon}$ determines how long the wave packet maintains its
acceleration. For $\ds{\epsilon<<1}$, accelerating behavior of the
ideal Airy wave packet was shown to be observed \cite{deney1}.\\
To summarize, we have presented self-accelerating matter waves for the harmonically
trapped condensate. We have derived formulas for self-acceleration and shown that the initial form of the wave function and an initial phase contribute the self-acceleration. We have discussed that nonlinear interaction has nothing to do with the self acceleration but changes the density profile of the self-accelerating wave. It is worth studying self-accelerating waves in an optical lattice potential.

\end{document}